\documentclass[aps,prb,floatfix,showpacs,twocolumn,groupedaddress]{revtex4}
\usepackage[english]{babel}
\usepackage{graphicx}
\usepackage{dcolumn}
\usepackage{amsmath}

\newcommand{\vs}{{\it vs.}}
\newcommand{\ie}{{\it i.e.}}
\newcommand{\al}{{\it et al.}}
\newcommand{\laysrca}{(La,Y)$_y$\-(Sr,Ca)$_{14-y}$\-Cu$_{24}$\-O$_{41}$}
\newcommand{\srca}{Sr$_{14-x}$\-Ca$_x$\-Cu$_{24}$\-O$_{41}$}
\newcommand{\sr}{Sr$_{14}$\-Cu$_{24}$\-O$_{41}$}
\newcommand{\lathree}{La$_3$\-Sr$_3$\-Ca$_8$\-Cu$_{24}$\-O$_{41}$}
\newcommand{\lafive}{La$_{5.2}$\-Ca$_{8.8}$\-Cu$_{24}$\-O$_{41}$}
\newcommand{\yzerofivefive}{Y$_{0.55}$\-Sr$_{13.45}$\-Cu$_{24}$\-O$_{41}$}

\newcommand{\yonesix}{Y$_{1.6}$\-Sr$_{12.4}$\-Cu$_{24}$\-O$_{41}$}
\newcommand{\ysr}{Y$_y$\-Sr$_{14-y}$\-Cu$_{24}$\-O$_{41}$}

\begin{document}

\title{Crossover from one-dimensional copper-oxygen chains to two-dimensional
ladders charge transport in \laysrca{}}

\author{T.\ Ivek}
\email{tivek@ifs.hr}
\homepage{http://real-science.ifs.hr/}
\author{T.\ Vuleti\'{c}}
\author{B.\ Korin-Hamzi\'{c}}
\author{O.\ Milat}
\author{S.\ Tomi\'{c}}
\affiliation{Institut za fiziku, P.O.Box 304, HR-10001 Zagreb, Croatia}
\author{B.\ Gorshunov}
\altaffiliation{Permanent address: A.\ M.\ Prokhorov General Physics Institute,
Russian Academy of Sciences, 119991 Moscow, Russia.}
\author{M.\ Dressel}
\affiliation{1. Physikalisches Institut, Universit\"{a}t Stuttgart, D-70550
Stuttgart, Germany}
\author{J.\ Akimitsu}
\author{Y.\ Sugiyama}
\affiliation{Department of Physics, Aoyama-Gakuin University, Setagaya-ku, Tokyo
157-8572, Japan }
\author{C.\ Hess}
\author{B.\ B\"{u}chner}
\affiliation{Leibniz-Institut f\"{u}r Festk\"{o}rper- und Werkstoffforschung,
D-01171 Dresden, Germany}
\date{\today}

\begin{abstract}
The charge transport in the copper-oxygen chain/ladder layers of \laysrca{} is
investigated along two crystallographic directions in the temperature range from
50~K to 700~K and for doping levels from $y \approx 6$ (number of holes
$n_h <1$) to $y = 0$ (number of holes $n_h = 6$). A crossover from a
one-dimensional hopping transport along the chains for $y \geq 3$ to a
quasi-two-dimensional charge conduction in the ladder planes for $y \lesssim 2$
is observed. This is attributed to a partial hole transfer from chains to
ladders when the hole doping exceeds $n_h \approx 4$ and approaches fully doped
value $n_h = 6$. For $y \lesssim 2$ a weak dielectric relaxation at
radio-frequencies and a microwave mode are detected, which might be recognized
as signatures of a charge-density wave phase developed at short length scales in
the ladders planes. 

\end{abstract}

\pacs{74.72.Jt, 71.27.+a, 72.20.Ee, 71.45.Lr}

%
%
%
%
%
%
%

\maketitle

\section{Introduction}
The spin-ladder and spin-chain systems \laysrca{} belong to a vast class of
strongly correlated materials, transition metal oxides, which exhibit some of
the most intriguing phenomena in condensed matter physics.\cite{Maekawa04} The
huge literature on the topics of spin-chains and spin-ladders accumulated in the
last decade (for a review see Ref.\ \onlinecite{Vuletic06}) has been triggered
by the discovery of superconductivity under pressure in the compound \srca{},
$x = 13.6$, mostly since this system is the first superconducting copper oxide
material with a non-square-lattice.\cite{Uehara96} The parent material, \sr{},
of this cuprate superconductor is a charge density wave (CDW) insulator with a
spin gap.\cite{Gorshunov02,Blumberg02,Abbamonte04,Kumagai97} Substituting
isovalent Ca for Sr suppresses the CDW insulating phase as shown by dc and ac
transport measurements,\cite{Vuletic03PRL} while the spin gap remains
constant.\cite{Kumagai97,Vuletic06} Nevertheless, recent resonant soft X-ray
scattering results \cite{Rusydi06} indicate that the CDW might stabilize even
for the highest Ca substitution but with a different periodicity indicating
strong commensurability effects. When for the compounds with high Ca content
external pressure is applied, the spin gap decreases in size, but remains finite
even when SC sets in.\cite{Piskunov01} Applying pressure also increases
interladder coupling leading to metallic transport along both the legs and rungs
of the ladders, \cite{Nagata98} as well as raises the number of mobile
quasi-particles at low temperature.\cite{Fujiwara03,Piskunov04} These particles
have a finite density of states at the Fermi level and might contribute to the
superconducting instability. All these results together with an indication for
the existence of a Hebel-Slichter coherence peak in the SC state as well as the
significant level of disorder in the doped ladders of \srca{}, indicate that the
superconducting pairing mechanism and symmetry are probably different compared
to theoretical predictions for pure single ladders.

By now it is well understood that the amount of doped holes and their
distribution between chains and ladders determines electronic phases and the
spin and charge dynamics. In the fully doped material \sr{} the total number of
holes ($n_h$) is six per formula unit. The hole distribution between chains and
ladders is probed most directly by the polarization-dependent near-edge X-ray
absorption fine structure (NEXAFS): at room temperature (RT) according to
N\"{u}cker \al{}\ \cite{Nuecker00} there is close to one hole per formula unit
transferred in the ladders (equivalent to $\delta = 0.07$ holes per ladder
copper site) and about five remain in the chains. Very recently, a quite
different distribution of close to three holes per formula unit on both ladders
and chains is suggested by Rusydi \al{}\cite{Rusydi07} The two-dimensional (2D)
ladders present a dominant charge transport channel: RT conductivity along the
$c$-axis is $\sigma_{\mathrm{dc}}(c) \approx 500$~$\Omega^{-1}$cm$^{-1}$ and
along the $a$-axis $\sigma_{\mathrm{dc}}(a) \approx 20$~$\Omega^{-1}$cm$^{-1}$.
Although $\sigma_{\mathrm{dc}}(c)$ is rather high, it shows an insulating
behavior, \ie{}\ it decreases with lowering temperature down to the
charge-density wave phase transition; a similar behavior is found in
$\sigma_{\mathrm{dc}}(a)$ as well.\cite{Vuletic05} At the same time, the
remaining holes in the chains negligibly contribute to the charge transport:
spin dimers are formed between those Cu$^{2+}$ spins that are separated by a
localized Zhang-Rice singlet (Cu$^{3+}$), that is, a site occupied by a
localized hole. In this way the antiferromagnetic (AF) dimer pattern is created
in chains together with the charge order, both inducing gaps in the spin and
charge sector,
respectively.\cite{Matsuda96,Matsuda96a,Takigawa98,Regnault99,note1}
On the other hand, in the underdoped materials \laysrca{} ($n_h = 6 - y$) the
absence of holes in ladders eliminates the ladder CDW phase, and suppresses the
charge-ordered gapped state in chains in favor of disorder-driven insulating
phase with charge transport by variable range hopping
(VRH).\cite{Vuletic03PRB,VuleticJPF05} The hopping transport originates in the
non-periodic potential in which holes reside and which is induced by strong
local distortions of the chains due to the irregular coordination of La$^{3+}$,
Y$^{3+}$, Sr$^{2+}$ and Ca$^{2+}$ ions. The VRH conductivity can then be
explained as a result of the distorted distribution of microscopic
conductivities, as predicted in Anderson's localization theory. In short, the
copper-oxygen chains in the underdoped quasi-1D cuprates can be considered like
a one-dimensional system in which disorder, associated with random distribution
of holes, causes the Anderson localization.

In our previous work \cite{Vuletic03PRB,Vuletic06,VuleticJPF05} we have
suggested that these results reveal an intriguing possibility for the existence
of a phase transition close to $n_h = 6$ in the phase diagram of \laysrca{}
compounds and that further experiments on materials with very low La/Y content,
which corresponds to $n_h \leq 6$, should elucidate our proposal. In this
article we attempt to answer this intriguing question on how and why and at
which doping level the one-dimensional hopping transport along the chains
crosses over into a quasi-two-dimensional charge conduction in the ladder
planes. In order to clarify this issue, we have undertaken dc and ac
conductivity-anisotropy measurements on single crystals of \laysrca{} with
different La/Y content (a particular emphasis was put on La/Y contents
approaching $y = 0$) in a wide frequency and temperature range. We show that for
the systems with $y \lesssim 2$ ($n_h \gtrsim 4$) variable-range hopping fails
as a relevant picture for the observed conductivity and that the charge sector
bears features encountered in the fully doped systems: conductivity anisotropy
is of similar order of magnitude and the logarithmic derivative of resistivity
presents (wide) maxima. These results suggest that in the underdoped systems
with doping levels $n_h \gtrsim 4$ ladders start contributing to charge
transport properties and prevail over chains as an electrical transport channel.
Concomitantly, frequency-dependent conductivity seems to indicate that charge
ordering at short scales starts to develop in the ladders.

\section{Sample characterization and experimental methods}
High-quality single crystals of materials with low Y content were synthesized:
$y = 0$ (\sr{}), $y = 0.55$ (\yzerofivefive{}), $y = 1.6$ (\yonesix{}). Samples
were characterized by powder X-ray diffraction and Y content was determined by
an electron probe microanalyzer. In this study needle-like samples of about
0.4~mm$^3$ in size were used, cut out of bulk single crystals of these
materials, together with previously synthesized $y = 3$ (\lathree{}) and
$y = 5.2$ (\lafive{}). Crystallographic orientation of crystals used in the
resistivity anisotropy measurements was determined by taking X-ray
back-reflection Laue patterns, and their subsequent simulation using
OrientExpress 3.3 software.\cite{Ognjensw} Simulated meshes of two overlapping
sublattice unit cells ($a = 11.47$~nm, $b = 13.37$~nm,
$c_\mathrm{L} = 3.93 \pm 0.03$~nm, Fmmm - for ladders; and $a = 11.47$~nm,
$b = 13.37$~nm, $c_\mathrm{C} = 2.73 \pm 0.03$~nm, Amma - for chains) fitted
well with the recorded patterns in case of proper crystallographic alignment.
The same crystallographic orientations were found in all single crystals: the
crystallographic $ac$ plane was found to be parallel to the largest faces of the
needle-like prismatic shape crystals: either $c$-axis or $a$-axis were properly
oriented lengthwise, along the needle axis. dc resistivity was measured between
50~K and 700~K. 

A Hewlett Packard 4284A and an Agilent 4294A impedance analyzers were used to
measure complex conductivities of $y = 0$, 0.55 and 1.6 at frequencies between
20~Hz and 10~MHz.\cite{Pinteric01} The data at the lowest frequency matched our
four-probe dc measurements. The complex dielectric function at frequencies
5--25~cm$^{-1}$ was obtained by complex transmission measurements using a
coherent-source THz spectrometer.\cite{BWO} For latter measurements crystals
with plane-parallel faces were prepared by polishing, with thickness of about
0.5~mm and of transverse dimensions about $7 \times 7$~mm$^2$. All measurements
were done along the two crystallographic axes defining chain and ladder layers:
$c$-axis (along the ladders legs and chains) and the $a$-axis (along the ladders
rungs).

\begin{figure}
\centering\includegraphics[clip,width=0.8\linewidth]{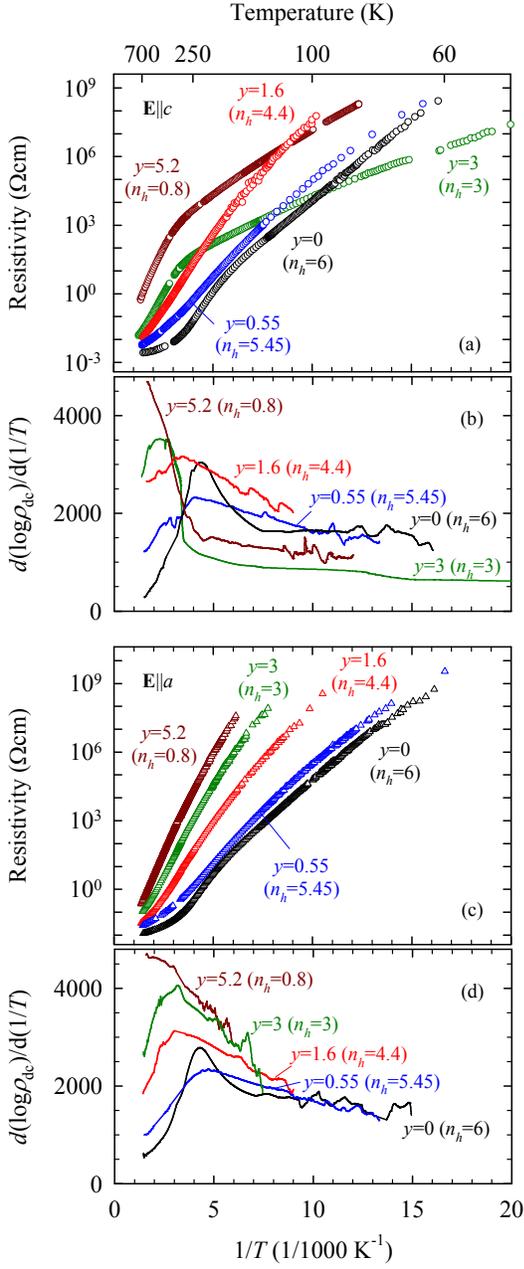}
\caption{(color online) dc resistivity and logarithmic derivatives of \laysrca{}
for various La/Y content $y$ along the $c$ [panels (a) and (b)] and the $a$
[panels (c) and (d)] crystallographic directions.}
\label{fig1}
\end{figure}

\section{Results and analysis}
\subsection{dc transport}

\begin{figure}
\centering\includegraphics[clip,width=0.8\linewidth]{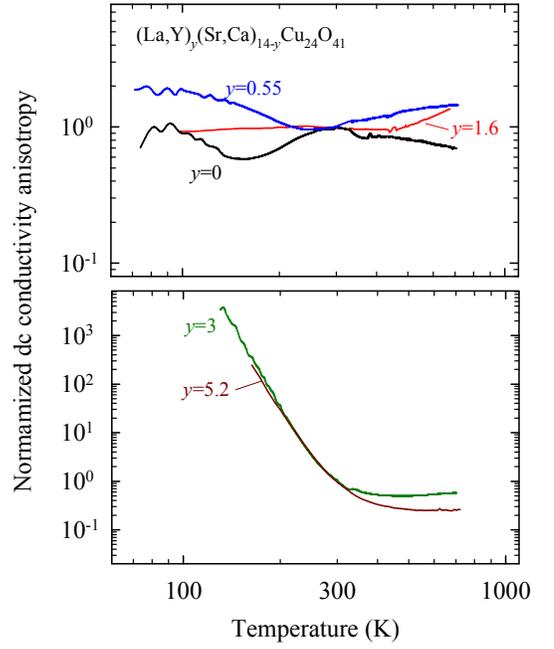}
\caption{(color online) Temperature dependence of the conductivity anisotropy of
\laysrca{} for various La/Y content $y$ normalized to the corresponding room
temperature value.}
\label{fig2}
\end{figure}

Fig.\ \ref{fig1} shows the behavior of dc resistivity and its logarithmic
derivative for different La/Y content ranging from $y = 5.2$ to $y = 0$ along
the $c$-axis [panels (a) and (b)] and the $a$-axis [panels (c) and (d)] in the
wide temperature range from 50~K (the lowest temperature obtained in our
experiment) up to 700~K. While for two compounds with high $y = 5.2$ and 3 the
dc resistivity curves along the $c$-axis and the $a$-axis markedly differ below
about 300~K, the one along the $c$-axis presenting a much smaller increase with
lowering temperature, one finds an almost identical behavior of dc resistivity
along the both axes for $y = 1.6$, 0.55 and 0. An immediate conclusion that can
be drawn from observed behaviors is that the conductivity anisotropy becomes
significantly enhanced for high La/Y content $y \geq 3$ (\ie{}\ low hole count
$n_h \leq 3$), whereas it remains small and temperature-independent for low $y$
(high $n_h$), as depicted in Fig.\ \ref{fig2}. The qualitative difference
between the two kinds of behavior is emphasized in Fig.\ \ref{fig2}, which shows
conductivity anisotropies normalized to the corresponding RT values. The
conductivity anisotropy at RT is in the range of 1-30 and basically does not
correlate with La/Y content.

The next significant difference between low and high La/Y contents is found in
the temperature dependence of the dc conductivity curves. As already reported
for $y = 5.2$ and 3, the dc conductivity along the $c$-axis
$\sigma_{\mathrm{dc}}(c)$ follows a variable-range hopping behavior with the
dimension of the system $d = 1$, and crosses over around $T_{\mathrm{co}}$ to
nearest neighbor hopping at high
temperatures.\cite{Vuletic03PRB,Vuletic06,VuleticJPF05} The observation of
$d = 1$ type of VRH conduction is in accord with a rather small interchain
coupling in \laysrca{}. Conversely, VRH fits
\begin{equation}
\sigma_{\mathrm{dc}}(T) = \sigma_0 \exp
\left[-\left(\frac{T_0}{T}\right)^{1/(1+d)}\right]
\label{eq1}
\end{equation}
to the $\sigma_{\mathrm{dc}}(c)$ curves for $y = 1.6$ and $y = 0.55$ fail to
give a meaningful description: the respective values of the VRH activation
energy $T_0^{\mathrm{exp}} = 13400$~meV and 9000~meV, obtained from the fit of
our data by expression (\ref{eq1}) are much larger than those for $y = 5.2$ and
3. This result is at variance with the behavior expected in the VRH mechanism:
the more conductive the sample, the lower $T_0$ is expected. Indeed, these
$T_0^{\mathrm{exp}}$ values are markedly different from the ones expected
theoretically: $T_0^{\mathrm{th}} = 2\cdot \Delta \cdot c_\mathrm{C} \cdot
\alpha \approx 1900$~meV and 700~meV, see Table \ref{table1}. Here the energy of
sites available for hops near the Fermi energy has a uniform distribution in the
range $-\Delta$ to $\Delta$, $c_\mathrm{C}$ is the distance between the nearest
Cu chain sites and $\alpha^{-1} = 2c_\mathrm{c} \cdot T_{\mathrm{co}} / \Delta$
is the localization length. In particular, the experimental values
$T_0^{\mathrm{exp}}$ are so high that the usual interpretation of the hopping
parameters also leads to values too low for the density of states for $y = 1.6$
and 0.55, when compared with $y = 5.2$ and 3. It can be noted that the
one-dimensional VRH conducting channel along the $c$-axis, which is present in
$y \geq 3$, is more efficient when compared with the transport in $y < 3$.

Comparing compounds with high and low $y$ the behavior of dc resistivity along
the $a$-axis, $\rho_{\mathrm{dc}}(a)$, differs in a manner that is qualitatively
alike $\rho_{\mathrm{dc}}(c)$. The slope of $\log \rho_{\mathrm{dc}}$
\vs{}\ $T^{-1}$ curves for $y=5.2$ and 3 shows that the activation energy is
much larger at high temperatures and becomes smaller with decreasing $T$,
whereas for $y = 0.55$ and 0 we find an opposite behavior: a smaller activation
energy at high temperatures and a larger one at low temperatures. It appears
that the $y = 1.6$ compound is situated somewhere at the border between these
two distinct behaviors. We recall that for $y = 0$ a smaller activation energy
at high temperatures and a larger at low temperatures is a feature associated
with an insulator-to-insulator phase transition into the CDW phase in the
ladders.\cite{Vuletic05}

\begin{table*}
\centering
\caption{dc transport parameters of \laysrca{} for various La/Y content $y$
along the $c$-axis.}
\begin{tabular*}{7.0in}{@{\extracolsep{\fill}}ccccccc}
\hline\hline
Compound & $y$ & $\Delta$~(meV) & $T_{\mathrm{co}}$~(K) &
$T_0^{\mathrm{exp}}$~(meV) & $\alpha^{-1}$~(\AA{}) & $T_0^{\mathrm{th}}$~(meV)\\
\hline
\yzerofivefive& 0.55 & $130\pm 40$ & $280\pm 15$ & $9000\pm 100$ & 0.960 & 750\\
\yonesix{}& 1.6 & $230\pm 10$ & $330\pm 30$ & $13400\pm 100$ & 0.677 & 1900\\
\lathree{}& 3 & $280\pm 10$ & $295\pm 5$ & $2500\pm 100$ & 0.481 & 3400\\
\lafive{}& 5.2 & $370\pm 50$ & $330\pm 5$ & $4300\pm 100$ & 0.435 & 4600\\
\hline\hline
\end{tabular*}
\label{table1}
\end{table*}

Another difference between compounds with low and high $y$ contents becomes
obvious when looking at the logarithmic derivative curves (Fig.\ \ref{fig1},
panels (b) and (d)). For $y=0.55$ (but not $y=3$ and 5.2), both $\mathbf{E}||a$
and $\mathbf{E}||c$ orientation show a broad and flat maximum in
$\mathrm{d}(\ln\rho)/\mathrm{d}(1/T)$ centred at about 210~K, similar to $y = 0$
where this feature, albeit more narrow, is recognized as a signature of the CDW
phase transition in the ladders. This feature remains visible for $y = 1.6$;
however it is now extremely broad and flat, shifted to 300~K and more pronounced
for $\mathbf{E}||a$ than in $\mathbf{E}||c$ orientation.

\begin{figure}
\centering\includegraphics[clip,width=1.0\linewidth]{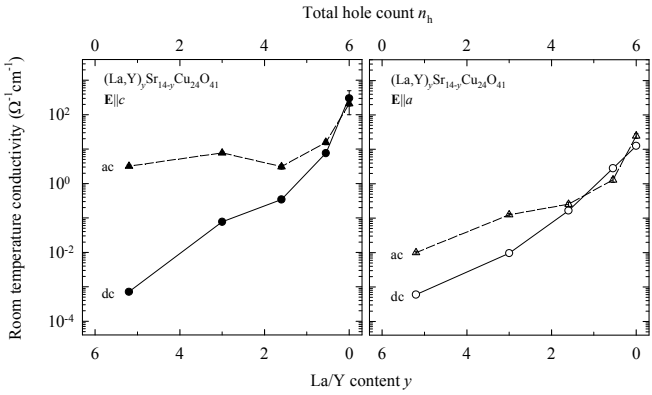}
\caption{Room temperature dc (circles) and ac conductivity at 10~cm$^{-1}$
(triangles) \cite{Borisnote} along the $c$-axis (left panel) and the $a$-axis
(right panel) as a function of La/Y content $y$ and total hole count $n_h$. The
full and dashed lines are guides for the eye for dc and ac data, respectively.}
\label{fig3}
\end{figure}

Finally, an unusual result concerns the magnitude of RT
conductivity\cite{Borisnote} along both axes which increases substantially with
total hole count (see Fig.\ \ref{fig3}). It is evident that the increased number
of holes per formula unit cannot account completely for this orders-of-magnitude
rise in conductivity. Theoretically, doping could create a finite density of
states at the Fermi level by shifting the Fermi level from the gap into the
region with high density of states, which then might partially account for the
observed conductivity rise.  Nevertheless, an overall rise hints to an
extraordinary increase of mobility which happens for $y$ smaller than two. 

\subsection{ac conductivity and dielectric function}
\begin{figure}
\centering\includegraphics[clip,width=1.0\linewidth]{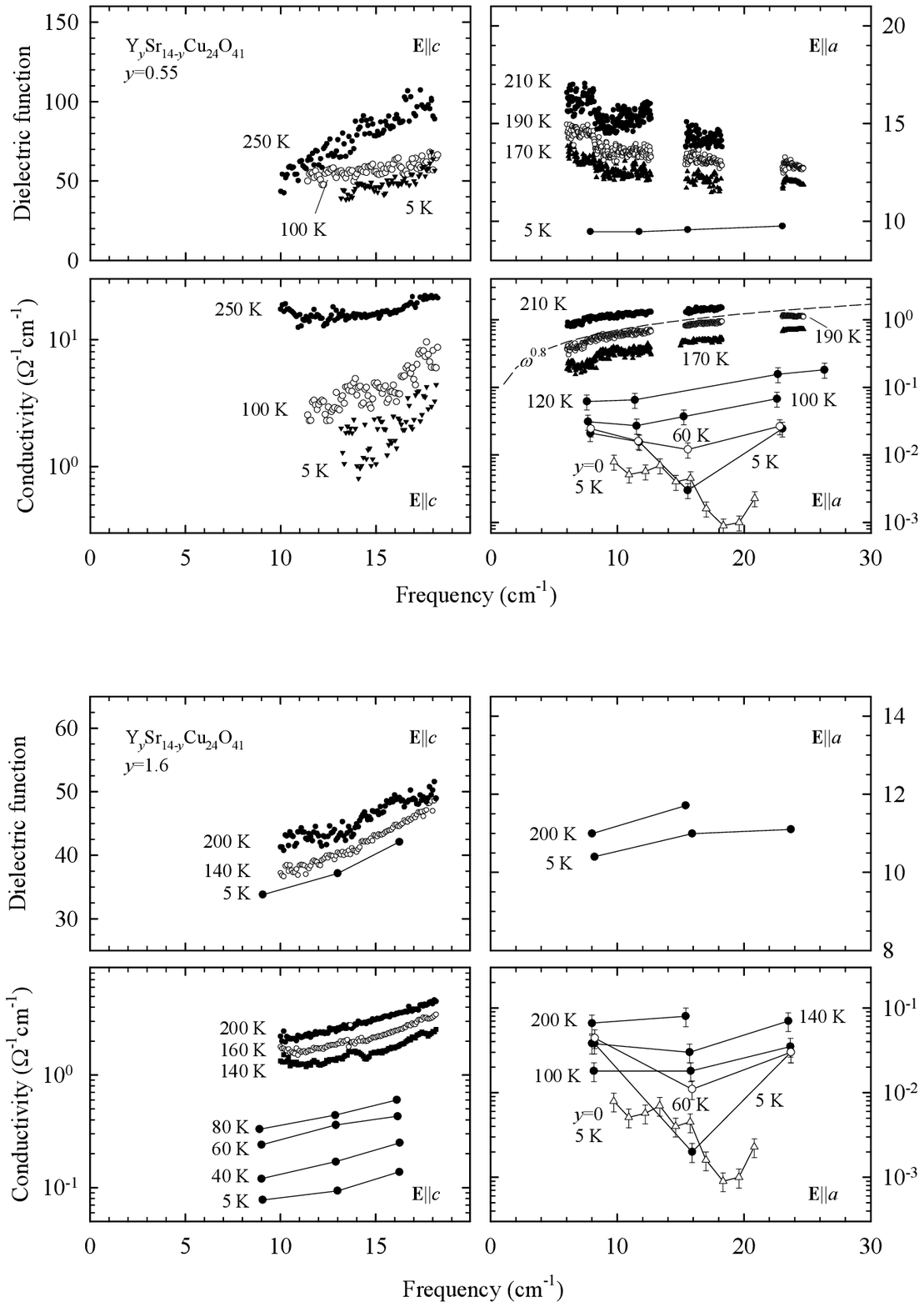}
\caption{Dielectric function and infrared conductivity in THz region of \ysr{}
for various Y content $y = 0.55$ (upper panel) and 1.6 (lower panel) along the
$c$- and $a$-axis at several temperatures as indicated. Conductivity data for
$y = 0$ along the $a$-axis at 5~K (denoted as open triangles) are shown for
comparison.}
\label{fig4}
\end{figure}

We now turn to other features derived from a comparative analysis of dc and ac
conductivity data. The conductivity spectra of \laysrca{} for $y = 0.55$ and 1.6
in the frequency range between 5 and 25~cm$^{-1}$ at several representative
temperatures are shown in Fig.\ \ref{fig4}. An almost dispersionless
conductivity spectrum at RT revealing the existence of a metallic response of
$y = 0$ in the infrared conductivity along both $c$-axis and $a$-axis (see
Fig.\ 67 in Ref.\ \onlinecite{Vuletic06}) is also evident for $y = 0.55$ and
1.6.\cite{note2} This result indicates the appearance of a certain amount of
free charges not detected for $y = 3$ and 5.2 (see Inset of Fig.\ 3 in
Ref.\ \onlinecite{Vuletic03PRB}) and indicates that the observed spectra could
be attributed to the charge excitations in the ladders similarly as for
$y = 0$.\cite{Vuletic06} We will address this behavior once more at the end of
this Section. On lowering the temperature below 200~K, a suppression of the
Drude weight is clearly visible in the conductivity spectra along the $c$-axis
and $a$-axis indicating that an insulating behavior develops.

\begin{figure}
\centering\includegraphics[clip,width=0.6\linewidth]{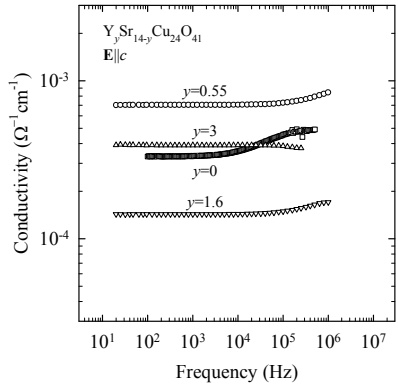}
\caption{Conductivity spectra of $y = 0$, 0.55, 1.6 and 3, $\mathbf{E}||c$ in
the radio-frequency range at representative temperatures (95~K, 125~K, 165~K and
132.5~K, respectively) with comparable dc conductivities.}
\label{fig5}
\end{figure}

\begin{figure}
\centering\includegraphics[clip,width=0.6\linewidth]{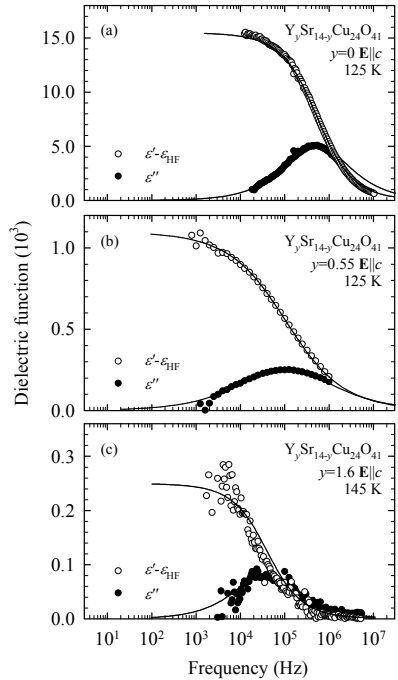}
\caption{Real ($\varepsilon^{\prime}$) and imaginary
($\varepsilon^{\prime\prime}$) parts of the dielectric function of \ysr{} for
$y = 0$ [panel (a)], $y = 0.55$ [panel (b)] and $y = 1.6$ [panel (c)] at
representative temperatures of 125~K ($y = 0$ and $y = 0.55$) and 145~K
($y = 1.6$) as a function of frequency, with the ac electric field applied along
the $c$-axis. The full lines are fits to data using the generalized Debye
expression $\varepsilon (\omega) - \varepsilon_{\mathrm{HF}} =
\Delta \varepsilon / ( 1 + ( i\omega \tau_0 )^{1 - \alpha} )$.}
\label{fig6}
\end{figure}

We note that in all studied cases (here we point out in particular $y = 0.55$
and $y = 1.6$) the conductivity behavior in the dc limit (see Fig.\ \ref{fig1})
is followed by a conductivity rise leading to a relatively high ac conductivity
in the infrared, as compared to the dc conductivity. A mechanism standardly
responsible for such a conductivity rise is electronic hopping conduction
characterized by a power-law dispersion $\sigma_{\mathrm{ac}}(T) \propto
A(T) \cdot \omega^s$. Indeed, hopping conduction with $s \approx 1$ has already
been established in the ladders of $y = 0$ compound for $\mathbf{E}||c$ and
$\mathbf{E}||a$, as well as in the chains of $y = 3$ and 5.2 for
$\mathbf{E}||c$.\cite{Vuletic03PRB,Vuletic06,VuleticJPF05} In this study, the
power-law behavior is found only for $y = 0.55$ ($\mathbf{E}||a$) between 200~K
and 100~K, freezing out at lower temperatures. There are two reasons which
prevented detection of hopping conduction for other cases. The first is related
to the phonon tail masking the hopping dispersion for $\mathbf{E}||c$
orientation.\cite{note3} Indeed, for the $c$-axis response of $y = 0.55$ and
1.6, at the lowest temperature ($T=5$~K) we see a typical phonon tail in the THz
range. It seems that for these two compositions the lowest frequency phonon sits
at about 25~cm$^{-1}$, \ie{}\ at the same frequency where the lowest frequency
phonon for the $y = 3$ (see Fig.\ 3 in Ref.\ \onlinecite{Vuletic03PRB}) and for
Sr$_{11}$Ca$_3$Cu$_{24}$O$_{41}$ compound was found. The second reason
preventing electronic hopping detection for $\mathbf{E}||a$ below about 100~K is
due to a clear conductivity increase below 20~cm$^{-1}$. We propose that this
increase might be an indication of a pinned CDW mode located in the microwave
range. It is noteworthy that this feature is also visible for $y = 0$ compound
(see Fig.\ \ref{fig4} for $\mathbf{E}||a$). Having only the higher frequency
slope of the mode we cannot make a quantitative fit and determine parameters
like eigenfrequency, dielectric strength and damping. Nevertheless, estimates
based on our dielectric function and conductivity data indicate that these
parameters would be much different from those of the pinned CDW mode at
1.8~cm$^{-1}$ as inferred by Kitano \al{}\cite{Kitano01} for fully doped
compound \sr{} based on some distinct microwave points, and as discussed at
length in Ref.\ \onlinecite{Vuletic06}. On the other hand, for $\mathbf{E}||c$
we do not detect any signature of this pinned mode in the THz range, which might
be either due to its location at lower frequencies, or the mode being masked by
a contribution of free carriers or a nearby phonon. It is noteworthy that this
mode, which we tentatively attribute to the pinned CDW mode, is absent in the
THz spectra of $y = 3$ and 5.2 compounds. This behavior indicates that an
alternative assignment of the CDW pinned mode emerging from our data, although
at delicate grounds due to a very narrow frequency range, might be of relevance
which should not be neglected. The issue of pinned CDW mode and its evolution in
\ysr{} obviously deserves more attention in the future. As far as dielectric
constant $\varepsilon^\prime$ of $y = 0.55$ and 1.6 is concerned, we note that
it coincides well with the dielectric constant of the fully doped compound
Sr$_{11}$Ca$_3$Cu$_{24}$O$_{41}$ (see Fig.\ 66 in Ref.\ \onlinecite{Vuletic06}),
meaning that the infrared phonon spectra of all these three materials could be
very similar. 

\begin{figure}
\centering\includegraphics[clip,width=0.6 \linewidth]{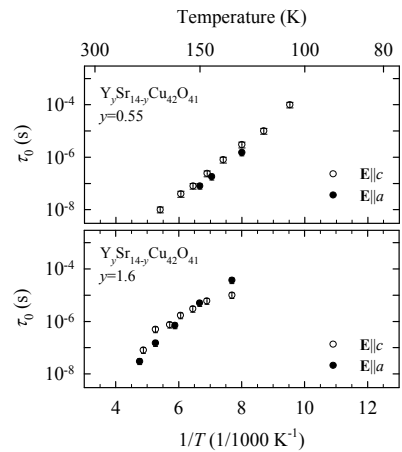}
\caption{Temperature dependence of mean relaxation times $\tau_0$ for $y = 0.55$
and 1.6. Open and full circles are for $\mathbf{E}||c$ and $\mathbf{E}||a$,
respectively.}
\label{fig7}
\end{figure}

We turn now to the radio-frequency results. While for $y = 0$ compound the CDW
develops in the ladders yielding a pronounced step-like conductivity increase in
the radio-frequency range, the frequency dependence is much weaker for
$y = 0.55$ and 1.6 and even comparable to $y = 3$ compound (see
Fig.\ \ref{fig5}). We recall that for $y = 3$ as well as for $y = 5.2$ the
frequency independent behavior is found in the radio-frequency range for all
temperatures.\cite{Vuletic03PRB,Vuletic06,VuleticJPF05} However, unlike for
$y = 3$ and y = 5.2, when the complex dielectric function for $y = 0.55$ and 1.6
is calculated from complex conductivity, a weak dielectric relaxation mode
emerges (see Fig.\ \ref{fig6}): a characteristic step-like drop in the real part
of dielectric function and a wide maximum in the imaginary part, resembling that
of the fully-doped \sr{} parent system ($y = 0$), where CDW is fully developed.
A similar behavior is observed for both polarizations $\mathbf{E}||c$ and
$\mathbf{E}||a$, as in the case of $y = 0$. Also, the mean relaxation time
$\tau_0$ has comparable values and temperature dependence when measured along
both the $c$- and $a$-axis (Fig.\ \ref{fig7}).\cite{Vuletic05,Vuletic06}
However, contrary to the $y = 0$ case, the temperature range in which we were
able to track the mode for $y = 0.55$ and 1.6 was rather narrow (see
Fig.\ \ref{fig7}). Still, a systematic trend in the behavior upon doping is
clearly visible. In this range the dielectric strength is small ($10^3$ and
$10^2$ for $y = 0.55$ and $1.6$, respectively) when compared to the value for
$y = 0$ ($10^4$ at same temperatures). Another worrisome issue is that, because
of the small low-frequency capacitance, we were not able to follow its
disappearance. Nevertheless, we are tempted to qualitatively associate this weak
mode with a ladder CDW order which persists only at short length scales for
$y = 0.55$ and 1.6, whereas it fully disappears for $y = 3$ and 5.2.

Finally, coming back to the crossover from metallic to insulating behavior upon
doping (\ie{}\ increasing $y$), we compare the RT ac conductivity at
10~cm$^{-1}$ with dc conductivity and find the following interesting feature
(see Fig.\ \ref{fig3}). RT conductivity data clearly show how the metallic-like
character of charge transport in $y = 0$ ($\sigma_{\mathrm{ac}}$ is close to
$\sigma_{\mathrm{dc}}$) gradually deteriorates with $y$ ($\sigma_{\mathrm{ac}}$
values differ from $\sigma_{\mathrm{dc}}$) and becomes typical for dielectrics
for $y = 3$ and 5.2. It is hard to quantify where this change starts since for
$y = 0.55$ and 1.6 the highest temperatures at which $\sigma_{\mathrm{ac}}$ was
measured were 210~K and 250~K, respectively,\cite{Borisnote} so that actual RT
$\sigma_{\mathrm{ac}}$ are higher than those shown in Fig.\ \ref{fig3}. Taking
this into account, the dc and ac conductivity contributions differ substantially
along both orientations for $y \gtrsim 2$, \ie{}\ when the total hole count is
smaller than four. At temperatures lower than about 200~K, reliable estimates of
infrared conductivity are prevented due to either a phonon or a pinned CDW-like
mode influence, so that we can only make crude estimates. However, we can say
that $\sigma_{\mathrm{ac}}(10$~cm$^{-1})/\sigma_{\mathrm{dc}}$ ratio for all
La/Y contents increases with lowering temperature, indicating the evolution of
the insulating behavior.

\section{Discussion}
From the above analysis we conclude that the one-dimensional hopping transport
along the chains for $2 < y \leq 6$ (hole-doping from zero to three injects
holes uniquely into chains) crosses over into a quasi-two-dimensional charge
conduction in ladders for smaller $y$. Supports for this conjecture are: a weak
and temperature-independent conductivity anisotropy (see Fig.\ \ref{fig2}) for
$0 \leq y \leq 1.6$; a maximum in $\mathrm{d}(\ln\rho)/\mathrm{d}(1/T)$ centered
around 210~K (see Fig.\ \ref{fig1}) which becomes broader and flatter going from
$y = 0$ to 1.6; a smaller activation energy at high temperatures and larger at
low temperatures: this difference disappears for $y = 1.6$. 

These results might be attributed to the CDW, whose long-range order (with
coherence length of about 260~\AA{}) as developed in $y = 0$
compound\cite{Abbamonte04} is destroyed, but domains developed at short range
scale still persist until $y \approx 1.6$. Indeed, a weak dielectric relaxation
mode is detected in the radio-frequency range; it resembles a CDW loss peak. The
increase of conductivity below 20~cm$^{-1}$ and the considerably larger value
compared to the dc conductivity infer some excitation which suggests an
additional mode somewhere in the microwave range. One might be tempted to
ascribe it to the pinned CDW mode, although the parameters seem to be different
from those of the peak that was proposed to be the pinned mode in $y=0$
compound.

Further, we remind that neutron scattering and static susceptibility
measurements show that with $y$ increasing from zero to one ($0 < y \leq 1$) AF
dimer long-range order in chains (AF dimers separated by a site occupied by a
localized hole) is also gradually destroyed.\cite{Matsuda97,Kato96,Herak} In
addition, NMR measurements of spin-lattice relaxation rate revealed that the
spin gap associated with AF dimer order in chains persists until
$y = 2$.\cite{Kumagai97} The latter result signals that antiferromagnetic and
charge correlations for $y = 2$ (total hole count $n_h = 4$) are already strong
enough that domains of AF dimers and related charge order (CO) form dynamically
and so exist at short time scales. Concomitantly, chains cease to be a favorable
charge transport channel and the beginning of hole transfer to ladders is
induced. A partial hole transfer from chains into ladders starts once the total
hole count becomes close to four and larger. Although probably only a tiny
amount of holes is transferred to ladders for $y = 1.6$, it appears that the
observed conduction with a weak and temperature-independent anisotropy happens
predominantly in ladders. For $y = 0.55$ it is evident that the charge transport
along the chains is almost completely frozen due to rather well developed AF
dimers and CO, and taken over by two-dimensional ladders in which transferred
amount of holes bears a much larger mobility, yielding an important conductivity
rise toward $y = 0$. 

Our results therefore suggest that ladders at La/Y content $y \lesssim 2$
prevail over chains as the conduction channel. A question arises why do holes,
which are doped only into the chains as La/Y content is varied from $y = 6$ to
$y \approx 2$, start to be distributed between chains and ladders once their
total count is larger than 4. In other words, it appears that doping more than 4
holes in the chains is energetically favorable only if at least a tiny amount of
holes is concomitantly doped in the ladders. A subtle interaction between chains
and ladders and stability of respective electronic phases in the charge and spin
sector is already evidenced for fully doped compounds: the chain CO and AF dimer
pattern on one side and ladder CDW on the other are both being suppressed at a
similar rate.\cite{Kataev01,Vuletic06} Our results in the underdoped series
follow on this idea and additionally reveal that the formation of these two
distinct electronic phases is also mutually interdependent, in the sense that
one cannot develop without the other.

As a final remark we note that the proposed scenario fits perfectly well to the
hole distribution proposed by N\"{u}cker \al{}\ \cite{Nuecker00} for $y = 0$
compound, \sr{}: close to 5 holes per formula unit in the chains and close to 1
hole per formula unit in the ladders. However, this hole distribution cannot
account for the observed periodicity of CDW in ladders.\cite{Abbamonte04}
Conversely, a hole distribution of close to three hole per formula unit in both
ladders and chains, recently proposed by Rusydi \al{},\cite{Rusydi07}
demonstrates opposite problems in explaining formation of electronic phases in
the underdoped series towards fully doped systems when La/Y content decreases
from $y = 3$ to $y = 0$, \ie{} when the total hole count increases from three to
six. Namely, a gradual doping of holes from zero to three in ladders nicely
explains formation of the CDW in ladders and its eventual periodicity found in
$y = 0$ compound when long-range order is developed. On the other hand, a fixed
hole count in chains in the range $0 \leq y < 3$, encounters difficulties to
explain short range AF dimer and CO domains therein, which dynamically appear at
$y \approx 2$ and grow in size as La/Y content decreases to $y = 0$. It also
stays in contradiction with the susceptibility results, which show that on
decreasing $y$ in the range $0 \leq y \leq 3$ the number of spins in chains
decreases, meaning a gradual increase of hole count in
chains.\cite{Kato96,Herak} Obviously, more experimental efforts are needed to
clarify and reconcile these contradictory findings in order to construct a
self-consistent picture of physics of chains and ladders in \laysrca{}.

\section{Conclusion}
In conclusion, we demonstrated the crossover from a one-dimensional hopping
charge transport in the chain subsystem for $y \geq 3$ to a
quasi-two-dimensional charge conduction in the ladder planes for $y \lesssim 2$.
We suggest that, while holes are doped only into the chains for low hole counts,
they are distributed between chains and ladders once the total hole count $n_h$
exceeds four. We propose that a clue which determines the hole distribution is
associated with a mutually interdependent formation of antiferromagnetic dimer
and charge order in chains and charge-density wave in ladders. Our results
confirm once more a profound interplay between chain and ladder sub-units,
showing clearly that any decent theoretical model attempting to give a proper
and self-consistent description of electronic phases in \laysrca{} should take
this into account.

\section*{Acknowledgements}
We thank G.\ Untereiner for the samples preparation, and M.\ Herak and
M.\ Miljak for useful discussions. This work was supported by the Croatian
Ministry of Science, Education and Sports under Grants No.035-0000000-2836 and
035-0352843-2844, the Deutsche Forschungsgemeinschaft (DFG), and by the Program
for Fundamental Research ``Problems of Radiophysics'' of the Department of
Physical Sciences, Russian Academy of Sciences. This work was partly supported
by the 21$^{\mathrm{st}}$ COE program, ``High-Tech Research Center'' Project for
Private Universities: matching fund subsidy from MEXT (Ministry of Education,
Culture, Sports, Science and Technology; 2002-2004), and a Grant-in-Aid for
Scientific Research on Priority Area from the Ministry of Education, Culture,
Sports, Science and Technology of Japan.


\begin{thebibliography}{99}
\bibitem{note1}
The main condition to establish AF dimer + CO is that the number of holes in the
chains per formula unit is close to six. This requirement is satisfied according
to Ref.\ \onlinecite{Nuecker00}, while it cannot be reconciled with the hole
distribution suggested in Ref.\ \onlinecite{Rusydi07}.
\bibitem{Borisnote} 
The measurement of ac conductivity for y = 0.55 and 1.6 was done at 250 K and
200 K, respectively, due to too high conductivity values causing too low
transmittivity of prepared plane-parallel samples.
\bibitem{note2}
While Drude response is clearly visible at RT for $y = 0.55$, $\mathbf{E}||c$
and $y = 1.6$, $\mathbf{E}||a$, it is masked by the phonon tail for $y = 1.6$,
$\mathbf{E}||c$.
\bibitem{note3}
We remind that only at low temperatures one standardly expects that due to
frozen charge carriers a low-energy phonon contribution prevails over electronic
hopping.
\end{thebibliography}
\end{document}